\pdfoutput=1
\documentclass[pra,preprint,showpacs,aps]{revtex4}

\usepackage{amsmath,amssymb}
\usepackage{latexsym}
\usepackage{graphicx}
\usepackage{epstopdf}
\usepackage{graphicx,amsfonts,amssymb}

\newcommand{\middlefig}{.45\textwidth}
\newcommand{\middleefig}{.35\textwidth}
\newcommand{\neff}{n_\mathrm{eff}}
\newcommand{\Dneff}{\Delta n_\mathrm{eff}}
\newcommand{\Dneffb}{\Delta n_\mathrm{eff,b}}
\newcommand{\Dneffd}{\Delta n_\mathrm{eff,d}}

\begin{document}
\title{Dark-bright gap solitons in coupled-mode one-dimensional\\ saturable waveguide arrays}

\author{Rong Dong, Christian E. R\"uter, Detlef Kip}
\affiliation{Faculty of Electrical Engineering, Helmut Schmidt University, 22043 Hamburg, Germany}
\author{Jes\'us Cuevas}
\affiliation{Grupo de F\'{\i}sica No Lineal, Departamento de F\'{\i}sica Aplicada I,
Escuela Polit\'ecnica Superior, C/ Virgen de \'{A}frica, 7, 41011
Sevilla, Spain}
\author{Panayotis G. Kevrekidis}
\affiliation{Department of Mathematics and Statistics, University of Massachusetts,
Amherst MA 01003-4515, USA}
\author{Daohong Song, Jingjun Xu}
\affiliation{The Key Laboratory of Weak-Light Nonlinear Photonics, Ministry of Education and TEDA, Applied Physics School,
Nankai University, Tianjin 300457, China}

\begin{abstract}
In the present work, we consider the dynamics of dark solitons as one
mode of a defocusing photorefractive lattice coupled with bright
solitons as a second mode of the lattice. Our investigation is motivated by
an experiment which illustrates that such coupled states can exist
with both components in the first gap of the linear band spectrum. This finding
is further extended by the examination of different possibilities from a theoretical
perspective, such as symbiotic ones where the bright component is
supported by states of the dark component in the first or second gap,
or non-symbiotic ones where the bright soliton is also a first-gap
state coupled to a first or second gap state of the dark component.
While the obtained states are generally unstable, these instabilities
typically bear fairly small growth rates which enable their
observation for experimentally relevant propagation distances.
\end{abstract}


\maketitle

\newpage

\section{Introduction}
The examination of the Hamiltonian continuum model with periodic potentials
and its discrete analog of lattice dynamical systems has been a topic of
increasing popularity over the past few years~\cite{reviews}. This is mainly
due to their wide applicability in diverse physical contexts including,
but not limited to, the spatial dynamics of optical beams in coupled waveguide arrays \cite{reviews1},
optically-induced photonic lattices in nonlinear optics \cite{reviews1a},
temporal evolution of Bose-Einstein condensates (BECs)
in optical lattices in soft-condensed matter physics \cite{reviews2}, and
the DNA double strand in biophysics \cite{reviews3}.

A principal research theme in this direction is
the study of existence and stability of coherent structures in these
models and their feasibility in experiments. Several years ago,
fabrication of nonlinear optical AlGaAs waveguide arrays \cite{7}
provided a first prototype where many initial investigations arose, such
as discrete diffraction, Peierls barriers, diffraction management~\cite{7a},
and gap solitons~\cite{7b}.  So far numerous fundamental
investigations have been pursued in waveguide arrays including
modulational instability \cite{MI_dnc}, four-wave-mixing effects arising
from the coupling of multiple components \cite{2c_dnc}, as well as the study
of interactions of solitary waves with surfaces \cite{dnc_surf}.
Subsequently, the formation of optically-induced photonic lattices in photorefractive
crystals became an ideal platform for the observation of various types of solitonic structures.
The theoretical proposal~\cite{solit} and rapid experimental realization of such (mainly 2D)
lattices~\cite{moti,moti2}, enabled the observation of, among others, dipole \cite{dip},
necklace \cite{neck}, and rotary \cite{rings} solitons as well as discrete \cite{vortex1,vortex2}
and gap \cite{motihigher} vortices.
Recently, waveguide arrays in lithium niobate (LiNbO$_3$) crystal,
which possess a self-defocusing nonlinearity, have found significant applications in the study
of modulation instability~\cite{MIchristian}, beam interactions~\cite{shand1}, dark discrete solitons~\cite{shand2},
bright gap solitons~\cite{brightSoliton1}, dark solitons in higher gaps~\cite{darksoliton1},
as well as Rabi oscillations \cite{Rabi}.

Our goal in this work is to consider the case of
vector solitons. Although they have been studied both
in the focusing case of bright-vector solitons in strontium barium niobate~\cite{zcv1}
and the defocusing case of bright-gap-vector solitons in LiNbO$_3$
~\cite{dkv1}, much less work has been done in multi-component settings.
Instead of mixtures of two solitary waves of the same type as in the above cases,
we aim to examine the mixture of a bright with a dark soliton in
photorefractive defocusing waveguide arrays.
Such dark-bright states were first created in the absence of lattices in  photorefractive
crystals over a decade ago~\cite{seg1}
and their interactions were partially monitored~\cite{seg2}.
In the context of BECs such solitary waves
were also predicted theoretically~\cite{BA}, and
generalizations thereof were considered as well (such as e.g.\ the
dark-dark-bright or bright-bright-dark
spinor variants of~\cite{DDB}). However, it was only quite
recently that such structures were
experimentally observed~\cite{hamburg,engels1,engels2,engels3}. This has led to
a renewed interest in this theme, by addressing the interactions of dark-bright solitons from an integrable
theory~\cite{rajendran} or numerical~\cite{berloff} perspective,
as well as their higher-dimensional generalizations~\cite{VB}.
To the best of our knowledge, there is no earlier investigation of
such states in models with a periodic potential except in the context of nonlinear dynamical
lattices~\cite{aj1}.

Our motivation, presented in section II stems from an
experiment in defocusing LiNbO$_3$ waveguide arrays where
a dark soliton state in the first gap
(we will refer to this type of state as ``bubble'' in what follows)
is coupled to a bright soliton in the same gap. We will show
that these two waveforms coexist as a solitonic entity. Also, we will
present conditions under which such a molecule may break
up in its constituents. This, in turn, motivates a more
detailed theoretical study of the different types of dark-bright states
that can exist in the system. Such coupled states will be identified between either a bubble (in the first gap)
or a higher-gap (i.e., the second gap in this case) dark soliton
in the one component with either a regular bright soliton or
with a bright gap soliton. When a bubble or dark soliton
couples to a regular bright one, we refer to these solitons as
symbiotic because the bright component can not exist without the supporting dark component (due to the defocusing
nature of the nonlinearity). For the coupling with a bright-gap soliton, because both components can persist individually,
we refer to these states as non-symbiotic. In section III,
we set up the model problem and benchmark it against experimental
data by identifying its linear band spectrum. In section IV,
the numerical results for the above soliton families will be given. Finally,
in section V we summarize our findings and present conclusions as well as some relevant directions for future study.

\section{Experimental Motivation}\label{sec:experimental}

To experimentally investigate such molecular solitonic states of dark and bright solitons, we used a 1D
waveguide array (WA) fabricated on an iron-doped lithium niobate (LiNbO$_{3}$) substrate by in-diffusion of titanium at high temperature. Arising from the bulk photovoltaic effect, the substrate crystal displays a saturable type of defocusing nonlinearity \cite{WA1}. The transverse direction $z$ is parallel to the ferroelectric $c$-axis. The direction of light propagation is along the $y$-axis. The array investigated in the following experiments consists of 250 channels and has a grating period $\Lambda=8.5\,\mu$m, which is the summation of the channel width of 5\,$\mu$m and a spacing of 3.5\,$\mu$m between adjacent channels. One of the end facets of the waveguide array sample is polished to optical quality to allow for direct observation of the out-coupled light from the array with the help of a CCD camera.

\begin{figure}
\begin{center}
    \includegraphics[width=80mm]{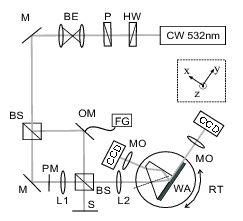}
\caption{Schematic experimental setup: HW, half-wave plate; P, polarizer; BE, beam expander; M's, mirrors; BS's, beam splitters; OM, oscillating mirror; FG, function generator; PM, phase mask; S, screen; L1 and L2, cylindrical lenses; MO's, microscopic objectives; CCD's, CCD cameras; PD, photodiode; WA, waveguide array.}
\label{Fig.1}
\end{center}
\end{figure}

In our experimental setup, we employed the prism-coupler scheme, with which
we can selectively excite different Bloch modes in any desired band. Furthermore, with this method we can determine accurately the band structure of
the waveguide array \cite{bandstructure1}. The experimental layout is sketched in Fig.~\ref{Fig.1}. First, the input light with a wavelength of 532\,nm from a frequency-doubled Nd:YVO$_{4}$ laser is expanded by a beam expander into a plane wave and then split into two separate beams. One beam propagates through a phase mask covering half of the beam along the transverse direction $z$. As a consequence, the covered half of the input beam experiences an additional $\pi$ phase shift, thus a dark notch is generated at the center of the intensity profile. Another beam is modulated by an oscillating mirror driven by a function generator. With applied external modulation, this beam is mutually incoherent with respect to the other beam. With the combination of two cylindrical lenses L1 and L2, the beam passing the phase mask is then imaged onto the waveguide. Here the focal lengths of the two lenses are chosen in order to generate an ideal width of the dark notch covering about two channels, which is the input light pattern for the excitation of a dark soliton. The other beam is focused meanwhile by lens L2 with a diameter of roughly 10\,$\mu$m and serves as the excitation light for the bright soliton. Both beams are coupled into the waveguide array and co-propagate until they reach the end-facet of the sample. With a high resolution CCD camera, in combination with a 20x microscopic objective lens, we can monitor around 25 channels of the intensity distribution on the end-facet. With this setup, it is possible to adjust the input light distribution for both, the bright soliton and the dark soliton separately, for example, the relative locations of the two solitons on the waveguide array as well as different excitation angles for modes originating from different bands.

\begin{figure}
\begin{center}
    \includegraphics[width=\middlefig]{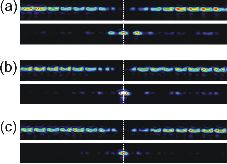}
\caption{Experimental results showing linear diffraction (a) and soliton
formation of individual bright and dark components (b). When both beams are
launched with a power ratio of 4:3, a robust bubble-bright soliton is formed (c) where both, bright and dark
components are centered on the same channel.}
\label{fig2}
\end{center}
\end{figure}

\begin{figure}
\begin{center}
    \includegraphics[width=\middlefig]{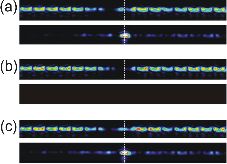}
\caption{When the input power ratio is increased (dominating bright
component) a shift of the dark soliton center is experimentally
observed (a). A restoration of the center position
of the bubble state appears when the bright component is blocked (b). This
process is reversible in the experiment, and the bubble is forced to shift
again from the center position when the bright beam is switched on again (c).}
\label{fig3}
\end{center}
\end{figure}

In the experiment, a bright gap soliton was excited from the first
and a dark soliton from the second band (a ``bubble''
according to our notation above), both at the edge of the Brillouin zone. The centers of both solitons were carefully adjusted to overlap on the same waveguide channel. We first checked under low optical power (less than 2\,nW per channel) the linear diffraction behavior of both the dark component [Fig.~\ref{fig2}(a), top row] and bright component [Fig.~\ref{fig2}(a), bottom row]. Then, by blocking one of the input beams, we formed individual gap solitons (either dark or bright) by increasing the optical power to appropriately high values [Fig.~\ref{fig2}(b)]. In all nonlinear experiments, the dark soliton from the second band was formed under 150\,nW optical power
per channel. In order to analyze the existence interval of the
bubble-bright
composite solitons, the input light power of the bright soliton
was varied, resulting in different power ratios of dark and bright components. At first, a bright soliton was formed at 200\,nW per channel, yielding a power ratio to the bright and dark solitons of 4:3. In this case, we observe a
robust co-existence of the two components at the output facet, as shown in
Fig.~\ref{fig2}(c).
%For these parameters, both the propagation constants of
%the coupled bright and dark components lie inside the existence region of
%dark-in-bright solitons, i.e.\ in the upper part of the numerically calculated $\mu$$_{d}$-$\mu$$_{b}$ plane in Fig. \ref{fig:existence}.
However, when we excite the bright soliton at much higher
power (400\,nW per channel, resulting in a power ratio of 8:3), the
propagation constant $\mu$$_{b}$ of the bright component in this scenario is
further decreased below the existence threshold (see also the theoretical
analysis below), while the propagation constant $\mu$$_{d}$ of the
dark component's bubble
state remains essentially unaffected. The result of the experiment in this
situation is a clear spatial shift of the bubble center by one waveguide
channel [Fig.~\ref{fig3}(a)] due to the coupling with the bright soliton.
This shift may be understood as the initial phase of a repulsive
interaction of the two constituents, and thus suggests the non-existence
(or strong instability) of bubble-bright solitons for these input
conditions. After reaching the steady-state for the input power ratio
8:3, we blocked the input beam used for excitation of the bright soliton.
Because the nonlinearity in lithium niobate is non-instantaneous, the
negative defect formed by the bright beam is still present and is only slowly
erased due to the photoconductivity generated by the remaining dark beam. As
a consequence, in the $\mu$$_{d}$-$\mu$$_{b}$ plane (see the left panel in Fig.~\ref{fig:existence}) we now move upwards
(i.e., $\mu$$_{b}$ increases) on a vertical line, reaching back
the existence regime of robust bubble solitary waves.
We thus observe a reversible effect, presented in Fig.~\ref{fig3}(b):
namely, after the bright soliton is blocked, the dark soliton is restored to
its original location. This restoration proves directly the repulsive
influence from the dominant bright soliton. When the bright component is switched on again in Fig.~\ref{fig3}(c),
%we again move to the non-existence region with
once again the strong repulsion between bright and dark components forces
the dark soliton to be shifted by one channel.\\

\section{Model setup}

In what follows, we will consider composite solitons with a dark
(or bubble) wave in one component coupled with a bright mode in
the second component in the context of TE-TE modes for the geometry of our waveguide
array. We start by
presenting the underlying model in the full dimensional form with the paraxial approximation, and then we discuss the non-dimensional
variant of the model which will be used for our numerical computations.

\subsection{Dynamical equations}

The paraxial equations for coupled TE-TE modes of the two
beams represented by $E_d$ and $E_b$ in what follows, are given by:

\begin{eqnarray}\label{eq:fulldyn}
    && i\partial_X E_d+\frac{1}{2k}\partial_{ZZ}E_d+
    \frac{k}{n_s}\left(\Delta n(Z)+\Delta n_{nl} \frac{|E_d|^2+|E_b|^2}{1+|E_d|^2+|E_b|^2}\right)E_d=0~, \nonumber \\
    && i\partial_X E_b+\frac{1}{2k}\partial_{ZZ}E_b+
    \frac{k}{n_s}\left(\Delta n(Z)+\Delta n_{nl} \frac{|E_d|^2+|E_b|^2}{1+|E_d|^2+|E_b|^2}\right)E_b=0~,
\end{eqnarray}

with $\Delta n(Z)$ being the refractive index profile and the propagation direction denoted as the $x$-direction. One can find ``stationary''
solutions of this system by defining:

\begin{equation}
    E_d(X,Z)=\mathrm{e}^{i\beta_d X}u(Z),\qquad
    E_b(X,Z)=\mathrm{e}^{i\beta_b X}v(Z),
\end{equation}

where $\beta_{d,b}$ are the propagation constants in the $X$-direction and $u(Z)$ and $v(Z)$ the amplitude profiles of each TE mode, which, in turn, satisfy:

\begin{eqnarray}\label{eq:fullstat}
    && -\beta_d u+\frac{1}{2k}\partial_{ZZ}u+
    \frac{k}{n_s}\left(\Delta n(Z)+\Delta n_{nl} \frac{u^2+v^2}{1+u^2+v^2}\right)u=0~, \nonumber \\
    && -\beta_b v+\frac{1}{2k}\partial_{ZZ}v+
    \frac{k}{n_s}\left(\Delta n(Z)+\Delta n_{nl} \frac{u^2+v^2}{1+u^2+v^2}\right)v=0~,
\end{eqnarray}

The values used in the experiments are the following ones:

\begin{equation}
    n_s=2.2341, \quad \lambda=532\ \textrm{nm},\quad \Lambda=8.5\ \mu\textrm{m}, \quad k=\frac{2\pi n_s}{\lambda}=26.386\ \mu\textrm{m}^{-1},
    \quad \Delta n_{nl}=2.5\times10^{-4}
\end{equation}
(cf. also the discussion given in section II) where $n_s$ is the refractive index of the LiNbO$_3$ substrate for extraordinary polarized light, $\lambda$ is the wavelength of the input light, $\Lambda$ is the period of the waveguide array and $\Delta n_{nl}$ is the maximum refractive index change induced by the nonlinearity.

The refractive index profile can be determined by adjusting the experimental
Bloch bands showing the change of the effective refractive
index $\Delta\neff\equiv\neff-n_s$, with $\neff=\beta k_0$ and $k_0=k/n_s$
being the transverse wavevector in vacuum. The refractive index is
then given by:

\begin{equation}
    \Delta n(Z)=\Delta n_0+\Delta n_1 V(Z)
\end{equation}

with

\begin{equation}
    V(Z)=\cos\left(\frac{2\pi Z}{\Lambda}\right)-0.25\cos\left(\frac{4\pi Z}{\Lambda}\right)
\end{equation}

and

\begin{equation}
    \Delta n_0-n_s=27.567\times10^{-4},\quad \Delta n_1=8.35\times10^{-4}~.
\end{equation}

\begin{figure}[hbf]
\begin{center}
    \includegraphics[width=\middlefig]{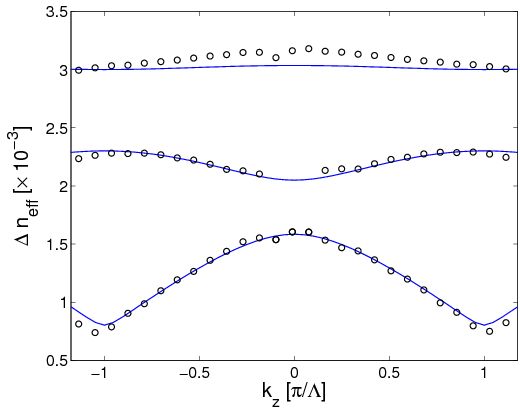}
\caption{Bloch bands numerically (full lines) and experimentally (circles) determined.}
\label{fig:Bloch}
\end{center}
\end{figure}

Figure \ref{fig:Bloch} shows the correspondence between the experimentally
observed Bloch bands~\cite{bandstructure1}  and the theoretically computed
ones. Clearly, the
above set of parameters offers a very good handle on the linear part
of the problem.

\subsection{Non-dimensional equations and parameters}

The non-dimensional version of the system of Eqs.
(\ref{eq:fulldyn}) is given by

\begin{eqnarray}\label{eq:dyn}
    && i\partial_x u+\frac{1}{2}\partial_{zz}u+[\eta^0+\eta V(z)]u+\nu \frac{u^2+v^2}{1+u^2+v^2}u=0~, \nonumber \\
    && i\partial_x v+\frac{1}{2}\partial_{zz}v+[\eta^0+\eta V(z)]v+\nu \frac{u^2+v^2}{1+u^2+v^2}v=0~,
\end{eqnarray}

while the stationary states are solutions of:

\begin{eqnarray}\label{eq:stat}
    && -\mu_d u+\frac{1}{2}\partial_{zz}u+[\eta^0+\eta V(z)]u+\nu \frac{u^2+v^2}{1+u^2+v^2}u=0~, \nonumber \\
    && -\mu_b v+\frac{1}{2}\partial_{zz}v+[\eta^0+\eta V(z)]v+\nu \frac{u^2+v^2}{1+u^2+v^2}v=0~.
\end{eqnarray}

The non-dimensional parameters are related to the experimental ones by the following relations:

\begin{equation}
    \mu_{d,b}=\frac{k\Lambda^2\beta_{d,b}}{\alpha^2},\qquad \nu=\pm\frac{k^2\Lambda^2\Delta n_{nl}}{\alpha^2 n_s}~,
\end{equation}

\begin{equation}
    \eta=\frac{k^2\Lambda^2\Delta n_{1}}{\alpha^2 n_s}, \qquad \eta^0=\frac{k^2\Lambda^2\Delta n_0}{\alpha^2 n_s}~.
\end{equation}

The parameter $\alpha$ has been introduced so that the non-dimensional values are of $O(1)$. Throughout the calculations, it has been fixed to $\alpha=10$. The sign of $\nu$ indicates either self-focusing (positive) or self-defocusing (negative). Additionally, the nondimensional distances are given by
\begin{equation}
    z=\alpha Z/\Lambda, \qquad x=\frac{\beta_{d,b}}{\mu_{d,b}}X=\frac{\alpha^2}{k\Lambda^2}X~.
\end{equation}

The refractive index profile and parameters are given now by

\begin{equation}
    V(z)=\cos\left(\frac{2\pi z}{\alpha}\right)-0.25\cos\left(\frac{4\pi z}{\alpha}\right)
\end{equation}

\begin{equation}
    \eta=0.1880,\qquad \eta^0=0.6207, \qquad \nu=\pm0.0563
\end{equation}

and the change of the effective refractive index is

\begin{equation}
    \Dneff=\frac{\beta_{d,b}\lambda}{2\pi}=\frac{\alpha^2\lambda^2\mu_{d,b}}{4\pi^2 n_s\Lambda^2}
\end{equation}

\noindent for each (dark and bright) component.

\subsection{Stability equations}

Once stationary solutions of the boundary value problem
(with periodic / anti-periodic boundary conditions, depending on
the nature of the examined solution) of
Eqs.~(\ref{eq:stat}) are identified, their linear stability is
considered by means of a Bogolyubov-de Gennes analysis. Namely, small
perturbations [of order ${\rm O}(\delta)$,
with $0< \delta \ll 1$] are introduced in the form

\begin{eqnarray}
    && E_d(z,x)=e^{i \mu_d x} \left[u_0(z) + \delta (P(z) e^{i \omega z} + Q^{*}(z) e^{-i \omega^{*} z}) \right]~, \nonumber \\
    && E_b(z,x)=e^{i \mu_b x} \left[v_0(z) + \delta (R(z) e^{i \omega z} + S^{*}(z) e^{-i \omega^{*} z}) \right]~,
\end{eqnarray}

and the ensuing linearized equation are then solved to O$(\delta)$, leading to the following eigenvalue problem:

\begin{equation}
    \omega
    \left(\begin{array}{c} P(z) \\ Q(z) \\ R(z) \\ S(z) \end{array}\right)=
    \left(\begin{array}{cccc}
        L_1 & L_2 & L_3 & L_3 \\ -L_2 & -L_1 & -L_3 & -L_3 \\
        L_3 & L_3 & L_4 & L_5 \\ -L_3 & -L_3 & -L_5 & -L_4
    \end{array}\right)
    \left(\begin{array}{c} P(z) \\ Q(z) \\ R(z) \\ S(z) \end{array}\right),
\end{equation}

for the eigenfrequency $\omega$ and the associated eigenvector $(P(z),Q(z),R(z),S(z))^T$, where
$L_j,\ j=1\ldots5$ are the following operators:

\begin{eqnarray}
    L_1 &\!=\!& -\mu_d+\frac{1}{2}\frac{d^2}{dz^2}+[\eta^0+\eta V(z)]+\nu
    \left[\frac{u_0^2+v_0^2}{1+u_0^2+v_0^2}+\frac{u_0^2}{(1+u_0^2+v_0^2)^2}\right]~,
\notag
\\[1.0ex]
    L_2 &\!=\!& \nu\frac{u_0^2}{(1+u_0^2+v_0^2)^2}~,
\notag
\\[1.0ex]
    L_3 &\!=\!& \nu\frac{u_0v_0}{(1+u_0^2+v_0^2)^2}~,
\notag
\\[1.0ex]
    L_4 &\!=\!& -\mu_b+\frac{1}{2}\frac{d^2}{dz^2}+[\eta^0+\eta V(z)]+\nu
    \left[\frac{u_0^2+v_0^2}{1+u_0^2+v_0^2}+\frac{v_0^2}{(1+u_0^2+v_0^2)^2}\right]~,
\notag
\\[1.0ex]
    L_5 &\!=\!& \nu\frac{v_0^2}{(1+u_0^2+v_0^2)^2}~,
\end{eqnarray}

where it has been taken into account that $u_0(z), v_0(z)\in\mathbb{R}$. Once the stationary solutions are found to be linearly unstable (i.e., ${\rm Im}\{\omega\} \ne 0$), then the dynamical manifestation of the corresponding instabilities is monitored through direct numerical simulations of Eq.~(\ref{eq:dyn}). As we will see in the next section, all of the analyzed solutions are unstable, although their growth rates are so small that long propagation distances $x$ are needed in order to observe the emergence of the pertinent instabilities.

\section{Numerical results}

We now present our results for the several types of coherent structures
considered in our system in the self-defocusing setting (i.e. $\nu<0$). All of
them are composed of a bright soliton in the 1st band gap. The dark
structure can be of two types. It may be a bubble, located in the 1st band
gap and arising from the top of the second Bloch band in which case the
overall phase shift between the two endpoints of the domain is $0$.
Alternatively, it may be
a (genuine) dark soliton, which emerges from the bottom of the second Bloch
band, and, consequently, its propagation constant is found in the second band
gap and it bears a phase shift of $\pi$ between the domain endpoints.

We make one more terminological distinction between the different types
of waveforms that can arise. In particular,
the emerging
bubble/dark-bright structures can either be symbiotic or not. In the first
case, the bright soliton is unstaggered and emerges from the top of the first
band (zero mode). These modes are called symbiotic because an isolated bright
component would {\it not} exist in this form for the relevant values of
the propagation constant; it necessitates the formation of an effective
potential by its dark (or bubble) counterpart in order to co-exist with it.
In the second (non-symbiotic) case, the bright soliton is staggered and
emerges from the bottom of the first band as a genuine gap soliton that
would be sustained in the system even in the absence of the other component.

\begin{figure}
\begin{center}
\begin{tabular}{cc}
    \includegraphics[width=\middleefig]{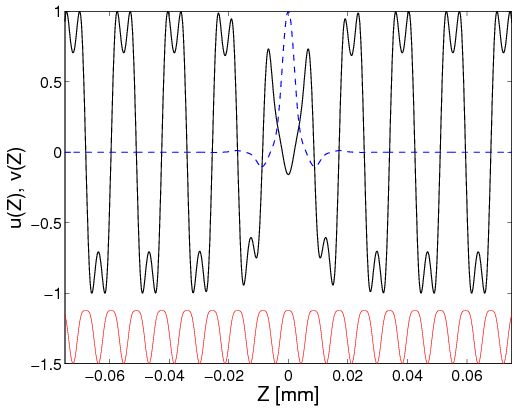} & \includegraphics[width=\middleefig]{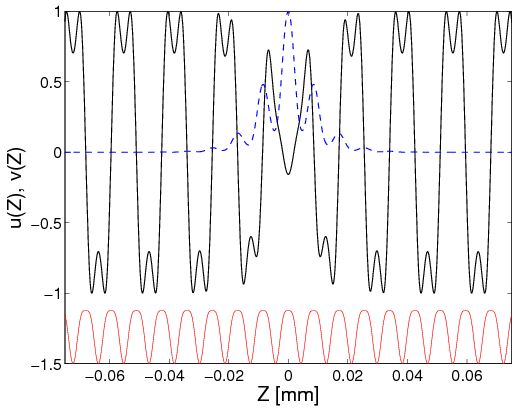}
\end{tabular}
\begin{tabular}{cc}
    \includegraphics[width=\middleefig]{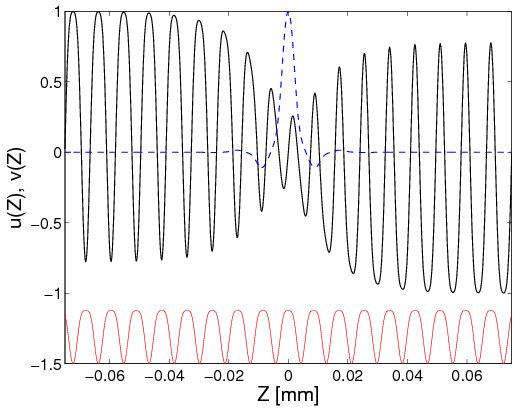} & \includegraphics[width=\middleefig]{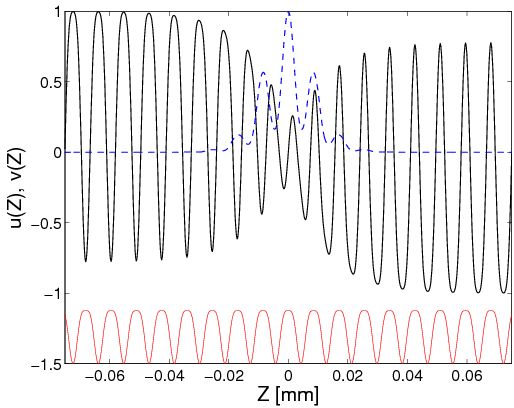}
\end{tabular}
%\begin{tabular}{cc}
 %   \includegraphics[width=\middleefig]{profile3a.jpg} & \includegraphics[width=\middleefig]{profile3b.jpg}
%\end{tabular}
%\begin{tabular}{cc}
%    \includegraphics[width=\middleefig]{profile4a.jpg} & \includegraphics[width=\middleefig]{profile4b.jpg}
%\end{tabular}
\caption{Profile of the electric field for a NSBBS with $\mu_b=0.64$ [$\Dneffb=2.84\times10^{-3}$] and $\mu_d=0.49$ [$\Dneffd=2.18\times10^{-3}$] (left panel in top line).
Profile of the electric field for a SBBS with $\mu_b=0.66$ [$\Dneffb=2.93\times10^{-3}$] and $\mu_d=0.48$ [$\Dneffd=2.13\times10^{-3}$] (right panel in top line).
Profile of the electric field for a NSDBS with $\mu_b=0.65$ [$\Dneffb=2.89\times10^{-3}$] and $\mu_d=0.33$ [$\Dneffd=1.47\times10^{-3}$] (left panel in bottom line).
Profile of the electric field for a SDBS with $\mu_b=0.67$ [$\Dneffb=2.98\times10^{-3}$] and $\mu_d=0.32$ [$\Dneffd=1.42\times10^{-3}$] (right panel in bottom line).
Blue dashed lines: input field of the bright component. Black solid lines: input field of the dark component. The red solid lines in each
case illustrate a rescaled form of $V(x)$ to indicate the location
of the potential wells.}
\label{fig:profiles1}
\end{center}
\end{figure}

These two distinctions (dark or bubble waves for the first component, symbiotic
or non-symbiotic ones depending on the nature of the second component) give
rise to four possibilities for the
ensuing structures dubbed as follows: symbiotic / non-symbiotic
bubble-bright soliton (SBBS / NSBBS) and symbiotic / non-symbiotic
dark-bright soliton (SDBS / NSDBS).
Among the four, it is the NSBBS that was observed in our experimental
motivation in section \ref{sec:experimental}.
Figure~\ref{fig:profiles1} %--\ref{fig:profiles4}
shows prototype examples of the input field profiles for each of these four solutions.

%\begin{figure}
%\begin{center}
%\begin{tabular}{cc}
%    \includegraphics[width=\middlefig]{profile2a.jpg} & \includegraphics[width=\middlefig]{profile2b.jpg}
%\end{tabular}
%\caption{Profile of the electric field for a SBBS with $\mu_b=0.66$ [$\Dneffb=0.002931324$] and $\mu_d=0.48$ [$\Dneffd=0.002131872$]. Red lines corresponds to a rescaled $V(x)$.}
%\label{fig:profiles2}
%\end{center}
%\end{figure}

%\begin{figure}
%\begin{center}
%\begin{tabular}{cc}
%    \includegraphics[width=\middlefig]{profile3a.jpg} & \includegraphics[width=\middlefig]{profile3b.jpg}
%\end{tabular}
%\caption{Profile of the electric field for a NSDBS with $\mu_b=0.65$ [$\Dneffb=0.00288691$] and $\mu_d=0.33$ [$\Dneffd=0.001465662$]. Red lines corresponds to a rescaled $V(x)$.}
%\label{fig:profiles3}
%\end{center}
%\end{figure}

%\begin{figure}
%\begin{center}
%\begin{tabular}{cc}
%    \includegraphics[width=\middlefig]{profile4a.jpg} & \includegraphics[width=\middlefig]{profile4b.jpg}
%\end{tabular}
%\caption{Profile of the electric field for a SDBS with $\mu_b=0.67$ [$\Dneffb=0.002975738$] and $\mu_d=0.32$ [$\Dneffd=0.001421248$]. Red lines corresponds to a rescaled $V(x)$.}
%\label{fig:profiles4}
%\end{center}
%\end{figure}

As mentioned above, Fig.~\ref{fig:Bloch} shows the position of the Bloch bands
which are also relevant for the identification of the nonlinear localized
modes that
arise in the system. In particular, the first band is located in the interval
$\mu_{1d}\equiv0.6755<\mu<0.6833\equiv\mu_{1u}$ [$3.00\times10^{-3}<\neff<3.03\times10^{-3}$], the second one is $\mu_{2d}\equiv0.4614<\mu<0.5181\equiv\mu_{2u}$ [$2.05\times10^{-3}<\neff<2.30\times10^{-3}$] and the third one at $\mu_{3d}\equiv0.1806<\mu<0.3567\equiv\mu_{3u}$ [$0.80\times10^{-3}<\neff<1.58\times10^{-3}$].

\begin{figure}
\begin{center}
\begin{tabular}{cc}
    \includegraphics[width=\middlefig]{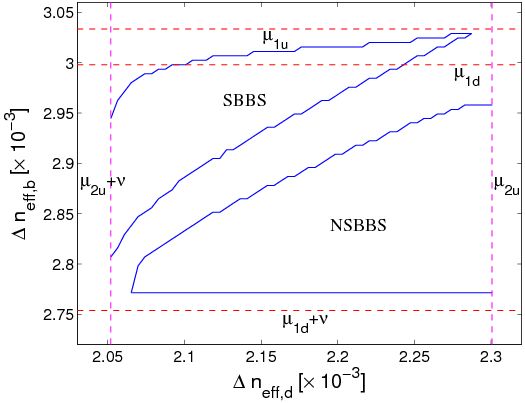} & \includegraphics[width=\middlefig]{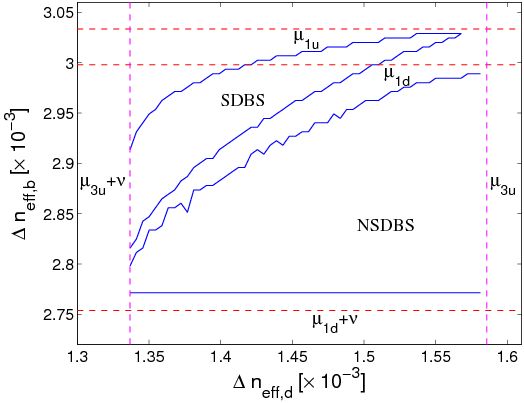}
\end{tabular}
\caption{Existence range for bubble-bright (left) and dark-bright (right) solitons. Relevant endpoints of the linear spectrum (and cutoff points below
which we were unable to continue the solution) are denoted by corresponding
horizontal or vertical dashed lines.}
\label{fig:existence}
\end{center}
\end{figure}

Our numerical computations show that, in absence of coupling between the
modes, the bright soliton can be identified in the first gap for
$\mu_{1d}+\nu<\mu_b<\mu_{1d}$ (i.e. $\mu_b\in[0.6192,0.6755]$, and
$\nu$ here as well as below denotes an appropriate shift) whereas
bubble-type solutions also exist for
$\mu_{2u}+\nu<\mu_d<\mu_{2u}$ (i.e. $\mu_d\in[0.4618,0.5181]$) and, in
turn, the dark soliton can be identified for lower values of the propagation
constant, namely for
$\mu_{3u}+\nu<\mu_d<\mu_{3u}$ (i.e. $\mu_d\in[0.3004,0.3567]$).
In the case of the two coupled beam components within the waveguide array,
the existence interval is narrower. Furthermore, the existence range depends
qualitatively
on the symbiotic / non-symbiotic character of the soliton. More specifically,
the accessible range of $\mu_d$, for a given $\mu_b$,
is always wider for symbiotic solitons than for non-symbiotic ones.
Additionally, the existence range of symbiotic solitons is limited from
above by $\mu_{1u}$. Figure~\ref{fig:existence} depicts the existence range
for dark-bright and bubble-bright symbiotic as well as
non-symbiotic solitary waves.
\begin{figure}
\begin{center}
\begin{tabular}{cc}
    \includegraphics[width=\middlefig]{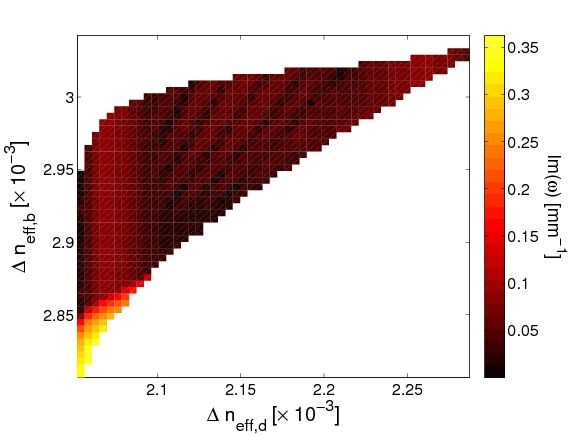} & \includegraphics[width=\middlefig]{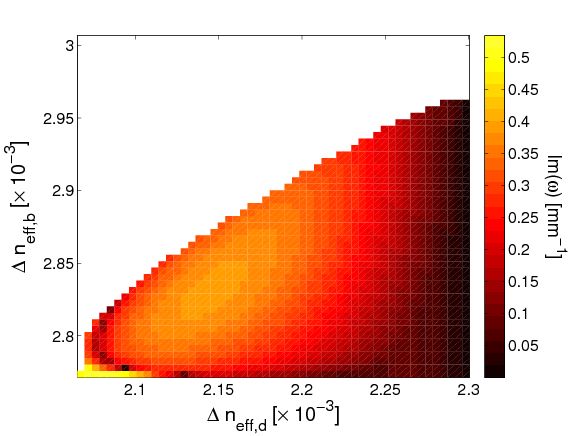} \\
    \includegraphics[width=\middlefig]{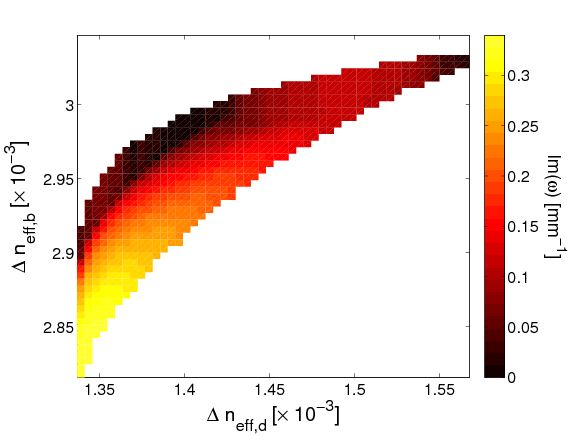} & \includegraphics[width=\middlefig]{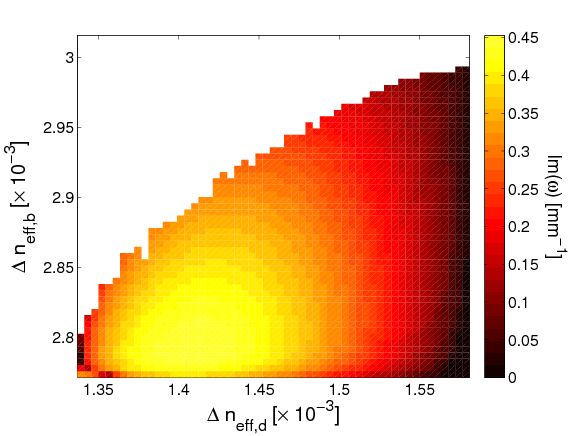}
\end{tabular}
\caption{Growth rates for bubble-bright (top) and dark-bright (bottom) solitons. Left (resp. right) panels correspond to symbiotic (resp.
non-symbiotic) structures.}
\label{fig:growth}
\end{center}
\end{figure}

\begin{figure}[ht]
\begin{center}
\begin{tabular}{cc}
    \includegraphics[width=\middleefig]{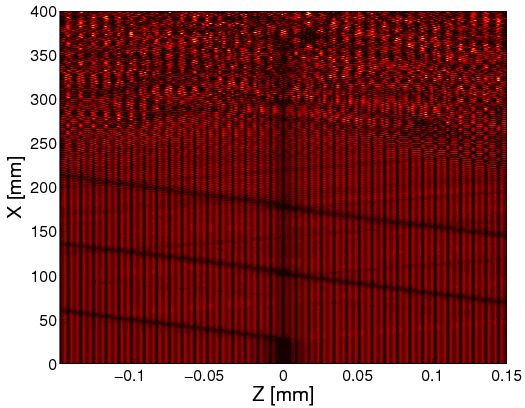} & \includegraphics[width=\middleefig]{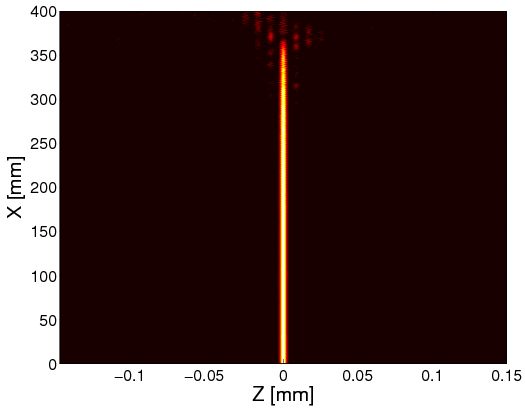} \\
    \includegraphics[width=\middleefig]{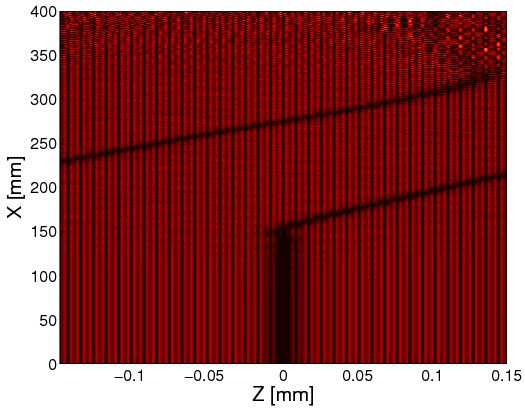} & \includegraphics[width=\middleefig]{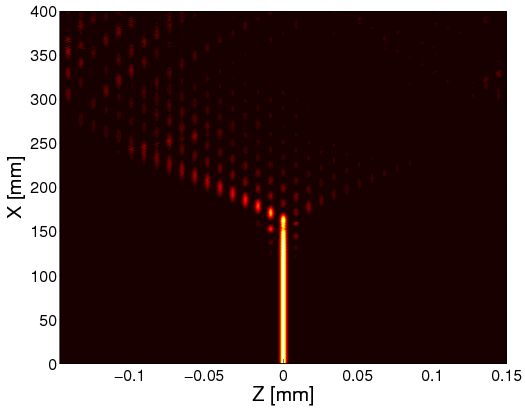} \\
    \includegraphics[width=\middleefig]{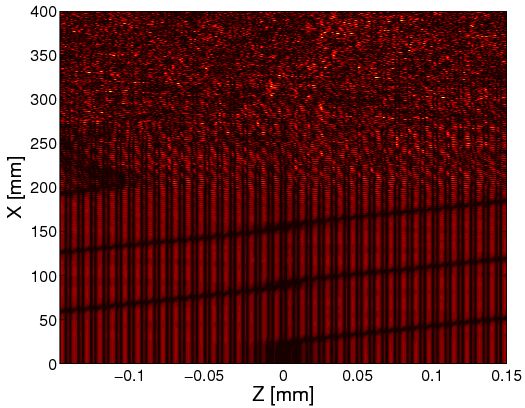} & \includegraphics[width=\middleefig]{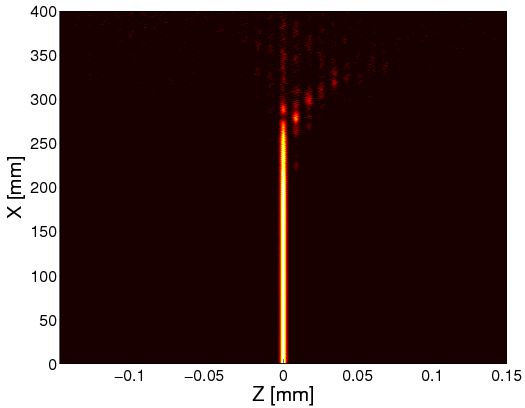} \\
    \includegraphics[width=\middleefig]{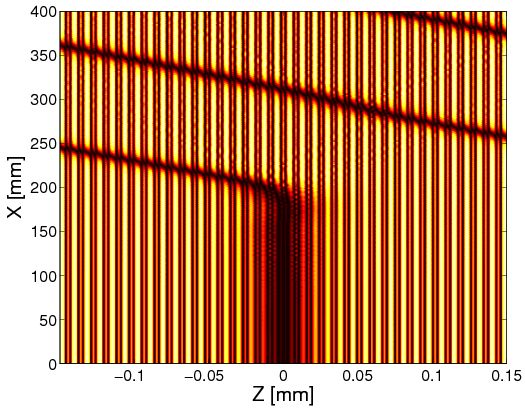} & \includegraphics[width=\middleefig]{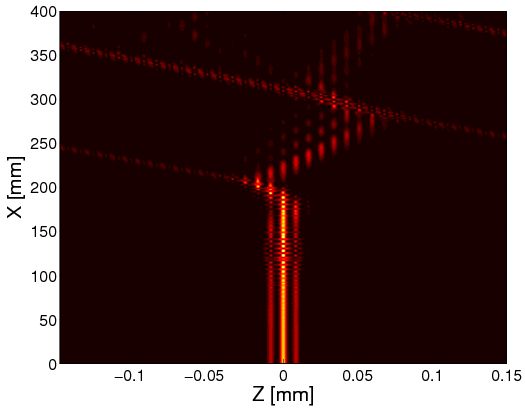}
\end{tabular}
\renewcommand{\baselinestretch}{0.55}
\caption{Top row: propagation of the squared modulus of the electric field for
the dark (left) and bright (right) components of a NSBBS with $\mu_d=0.48$ [$\Dneffd=2.13\times10^{-3}$] and $\mu_b=0.635$ [$\Dneffb=2.82\times10^{-3}$].
Second row: propagation of the squared electric field for the dark (left) and bright (right) components of a SBBS with $\mu_d=0.48$ [$\Dneffd=2.13\times10^{-3}$] and $\mu_b=0.67$ [$\Dneffb=2.98\times10^{-3}$].
Third row: propagation of the squared modulus of the electric field for the dark (left) and bright (right) components of a NSDBS with $\mu_d=0.32$ [$\Dneffd=1.42\times10^{-3}$] and $\mu_b=0.64$ [$\Dneffb=2.84\times10^{-3}$].
Bottom row:  propagation of the squared electric field for the dark (left) and bright (right) components of a SDBS with $\mu_d=0.31$ [$\Dneffd=1.38\times10^{-3}$] and $\mu_b=0.66$ [$\Dneffb=2.93\times10^{-3}$].}
%Bottom panels show the squared electric field for a propagation length $X=200$ mm.}
\label{fig:example1}
\end{center}
\end{figure}

\begin{figure}[ht]
\begin{center}
\begin{tabular}{cc}
    \includegraphics[width=\middlefig]{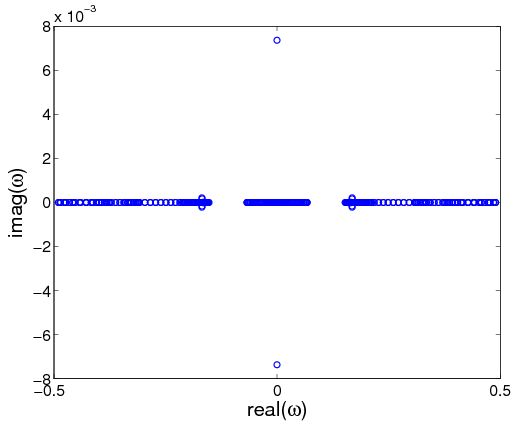} & \includegraphics[width=\middlefig]{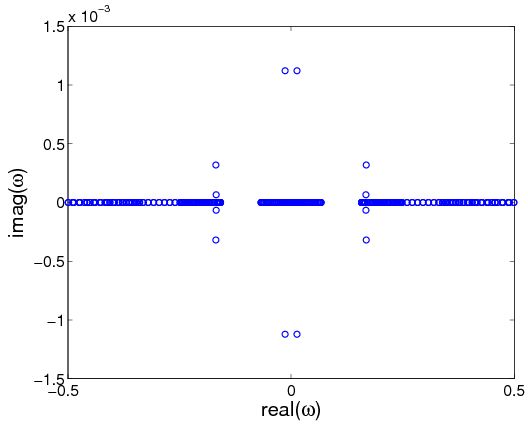} \\
    \includegraphics[width=\middlefig]{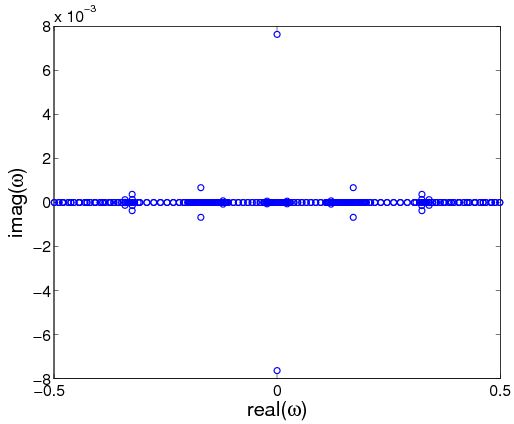} & \includegraphics[width=\middlefig]{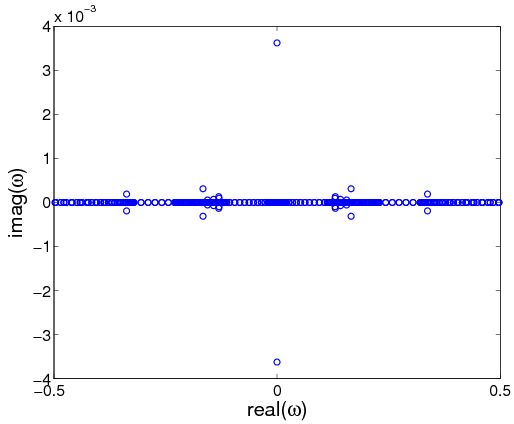}
\end{tabular}
\caption{Stability panels for the solitons of Fig. \ref{fig:example1},
namely for the NSBBS in the top left, the  SBBS in the top right,
the NSDBS of the bottom left and SDBS of the bottom right.}
%(top, left), Fig. \ref{fig:example2} (top, right), Fig. \ref{fig:example3} (bottom, left) and Fig. \ref{fig:example4} (bottom, right).}
\label{fig:spectra}
\end{center}
\end{figure}

We have examined the linear stability of the obtained solutions,
finding that the relevant waveforms are {\it generically} unstable in the
spectral sense. I.e., we have identified an imaginary or complex
eigenfrequency associated with the linearization spectrum around
these profiles, however
the growth rate is typically fairly small ($\lesssim 10^{-3}$
in non-dimensional units, i.e., $\lesssim 0.05$ mm$^{-1}$ in dimensional
units) and always
less than $10^{-2}$ in non-dimensional units
corresponding to 0.5 mm$^{-1}$ in dimensional ones.
Consequently, instabilities appear at a sufficiently large
propagation distance $X$ (inversely proportional to the above growth rate).
Figure~\ref{fig:growth} shows the growth rate dependence with $\Dneffd$ and
$\Dneffb$ for the four analyzed structures; notice the colorbar on the right
indicating the magnitude of the respective growth rates.
In order to test the effect of instabilities, a random perturbation of magnitude
$\sim10^{-3}$ is introduced to the input field profile. The main dynamical
observed outcome is the mobility of the dark component of the soliton.
This implies a break-up of the structure; however,
there are two realizations thereof depending on the symbiotic or
non-symbiotic nature of the state. In the case of a non-symbiotic solitons
(i.e., for NSBBS and for NSDBS),
the bright component remains at rest forming a genuine bright gap
soliton. On the other hand, in the symbiotic solitons, this is impossible
due to the non-existence of a bright waveform of this type.
Hence, most of the bright component energy moves towards the opposite
direction of the dark component in the case of the bubble (i.e., for SBBS)
while part of the energy moves with the dark component. For the
SDBS, most of the energy appears to move together with the dark component.
A summary of this scenario is shown in the panels of Fig.~\ref{fig:example1}.
To indicate the growth rates and unstable eigenmodes of the solutions
dynamically followed in Fig.~\ref{fig:example1}, we show in
Fig.~\ref{fig:spectra} their respective spectral planes.
It is worth remarking that, in most cases, the instabilities are of
exponential and oscillatory type, except in the case of SBBS, where most of
the instabilities are purely oscillatory. On the other hand,
to connect these results with the experimental motivation of
Section \ref{sec:experimental}, let us point out that for the
NSBBS considered therein the increase of the power is tantamount
to a larger instability growth rate and hence the observation of
the mobility of the dark component, while the bright one forms
a genuine gap soliton in agreement with our numerics  (top panel
of Fig.~\ref{fig:example1}). This repulsive effect between the
two components is also evident through the blocking of the bright
channel and the restoration of the bubble at the center, while
the reintroduction of the interaction between the beams naturally
and reversibly reinstates the repulsive bubble mobility effect.

\section{Conclusions and Future Challenges}

In the work presented in this paper, we have considered the case of two-component
dark-bright type solitary wave states in defocusing photorefractive
waveguide arrays. Motivated by experiments in LiNbO$_3$ arrays,
which illustrated a bubble-type soliton state in one
component coupled to a bright gap solitary wave in the second one,
we delved into a theoretical examination of the different composite
states that can emerge in this system. In particular, we revealed
the potential for four distinct types of waves, namely non-symbiotic
and symbiotic, dark-bright and bubble-bright ones. We numerically
revealed (within a model benchmarked against the linear band structure)
the persistence boundaries of such solutions. We also analyzed their
linear stablity which exhibits a typically weak instability in all of
them (with fairly small growth rates). This instability is so weak that
it permits, apparently, the experimental observability of the states.
Nevertheless, in suitable regimes even the experimental dynamics
manifests the potential break-up of the composite states.

Naturally, this investigation paves the way for numerous additional
studies. On the one hand, from an experimental viewpoint it would certainly
be interesting to identify the other proposed structures.
On the other hand, investigating interactions of such composite structures
would also offer relevant insights
as was done experimentally e.g.\ with simpler states in~\cite{shand1},
or as was done numerically in~\cite{aj1} and in different (BEC)
dark-bright contexts in~\cite{engels3,berloff}. Generalizations of
such states in two-dimensional waveguide arrays with the formation
of vortex-bright states~\cite{VB} or of genuinely discrete variants
thereof~\cite{interlaced} would also be an exciting theme for future
investigations.

\acknowledgments DK thanks the German Research Foundation (DFG, grant KI482/11-2)
for financial support of this research. PGK gratefully acknowledges support from
the US-NSF through grants DMS-0806762 and CMMI-1000337 and
from the Alexander von Humboldt Foundation as well as the Alexander
S. Onassis Public Benefit Foundation. JC acknowledges financial support from the MICINN project FIS2008-04848.


\newpage
\begin{thebibliography}{99}


\bibitem{reviews} S. Aubry,
Physica D {\bf 103}, 201, (1997);
S. Flach and C.R. Willis,
Phys. Rep. {\bf 295}, 181 (1998);
D. Hennig and G. Tsironis,
Phys. Rep. {\bf 307}, 333 (1999);
P.G. Kevrekidis, K.O. Rasmussen, and A.R. Bishop,
Int. J. Mod. Phys. B {\bf 15}, 2833 (2001);
A. Gorbach and S. Flach,
Phys. Rep. {\bf 467}, 1 (2008).

\bibitem{reviews1} D.N. Christodoulides, F. Lederer and Y. Silberberg,
Nature \textbf{424}, 817 (2003);
Yu.S. Kivshar and G.P. Agrawal,
\textit{Optical Solitons: From Fibers to Photonic Crystals}, Academic Press (San Diego, 2003).
%F. Lederer, G.I. Stegeman, D.N. Christodoulides, G. Assanto,
%M. Segev and Y. Silberberg,
%Phys. Rep. {\bf 463}, 1 (2008).

\bibitem{reviews1a} F. Lederer, G.I. Stegeman, D.N. Christodoulides,
G. Assanto, M. Segev, and Y. Silberberg, Phys. Rep. {\bf 463}, 1 (2008).



\bibitem{reviews2} P.G. Kevrekidis and D.J. Frantzeskakis,
Mod. Phys. Lett. B {\bf 18}, 173 (2004);
V.V. Konotop and V.A. Brazhnyi,
Mod. Phys. Lett. B {\bf 18}, 627, (2004);
P.G. Kevrekidis, R. Carretero-Gonz{\'a}lez, D.J. Frantzeskakis, and I.G. Kevrekidis,
Mod. Phys. Lett. B {\bf 18}, 1481 (2004).

\bibitem{reviews3} M. Peyrard, Nonlinearity {\bf 17}, R1 (2004).

\bibitem{7} H.S. Eisenberg, Y. Silberberg, R. Morandotti, A.R. Boyd,
and J.S. Aitchison, Phys. Rev. Lett. {\bf 81}, 3383 (1998).

\bibitem{7a} R. Morandotti, U. Peschel, J.S. Aitchison, H.S. Eisenberg, and
Y. Silberberg,  Phys. Rev. Lett. {\bf 83}, 2726 (1999).

\bibitem{7b} D. Mandelik, R. Morandotti, J.S. Aitchison, and Y. Silberberg,
Phys. Rev. Lett. {\bf 92}, 093904 (2004).

\bibitem{MI_dnc} J. Meier, G.I. Stegeman, D.N. Christodoulides, Y. Silberberg, R. Morandotti, H. Yang, G. Salamo, M. Sorel, and J.S. Aitchison,
 Phys. Rev. Lett. {\bf 92}, 163902 (2004).

\bibitem{2c_dnc} J. Meier, J. Hudock, D.N. Christodoulides, G. Stegeman, Y. Silberberg, R. Morandotti, and J.S. Aitchison,
Phys. Rev. Lett. {\bf 91}, 143907 (2003).

\bibitem{dnc_surf} S. Suntsov, K.G. Makris, D.N. Christodoulides, G.I. Stegeman, A. Hach{\'e}, R. Morandotti, H. Yang, G. Salamo, and M. Sorel,
Phys. Rev. Lett. {\bf 96}, 063901 (2006).

\bibitem{solit} N.K. Efremidis,
S. Sears, D.N. Christodoulides, J.W. Fleischer, and M. Segev,
Phys. Rev. E {\bf 66}, 046602 (2002).

\bibitem{moti} J.W. Fleischer, M. Segev, N.K. Efremidis, and D.N. Christodoulides,
%"Observation of two-dimensional discrete solitons in optically induced nonlinear photonic lattices''
Nature {\bf 422}, 147-150 (2003).

\bibitem{moti2} H. Martin, E.D. Eugenieva, Z. Chen, and D.N. Christodoulides,
Phys. Rev. Lett. {\bf 92}, 123902 (2004);
J.W. Fleischer, T. Carmon, M. Segev, N.K. Efremidis, and D.N. Christodoulides,
%"Observation of Discrete Solitons in Optically Induced Real Time Waveguide Arrays'',
Phys. Rev. Lett. {\bf 90}, 023902 (2003).

\bibitem{dip} J. Yang, I. Makasyuk, A. Bezryadina, and Z. Chen,
%"Dipole solitons in optically induced two-dimensional photonic lattices''
Opt. Lett. \textbf{29}, 1662-1664 (2004).

%\bibitem{quad} J. Yang, I. Makasyuk, A. Bezryadina, and Z. Chen,
%``Dipole and quadrupole solitons in optically induced two-dimensional photonic lattices: theory and experiment'',
%Stud. Appl. Math. \textbf{113}, 389-412 (2004).

\bibitem{neck} J. Yang, I. Makasyuk, P.G. Kevrekidis, H. Martin, B.A. Malomed, D.J. Frantzeskakis, and Z. Chen,
%``Necklacelike Solitons in Optically Induced Photonic Lattices'',
Phys. Rev. Lett. \textbf{94}, 113902 (2005).

%\bibitem{multi} Z. Chen, H. Martin, E.D. Eugenieva, J. Xu, and A. Bezryadina,
%Phys. Rev. Lett. \textbf{92}, 143902 (2004).

%\bibitem{fedele} F. Fedele, J. Yang, and Z. Chen,
%"Defect modes in one-dimensional photonic lattices'',
%Opt. Lett. {\bf 30}, 1506-1508 (2005).

\bibitem{rings}
Y.V. Kartashov, V.A. Vysloukh, and L. Torner, Phys. Rev. Lett.
{\bf 93}, 093904 (2004); X. Wang, Z. Chen, and P.G. Kevrekidis,
%"Observation of Discrete Solitons and Soliton Rotation in Optically Induced Periodic Ring Lattices'',
Phys. Rev. Lett. \textbf{96}, 083904 (2006).

\bibitem{vortex1} D.N. Neshev, T.J. Alexander, E.A. Ostrovskaya, Yu.S. Kivshar, H. Martin, I. Makasyuk, and Z. Chen,
% "Observation of Discrete Vortex Solitons in Optically Induced Photonic Lattices'',
Phys. Rev. Lett. {\bf 92}, 123903 (2004).

\bibitem{vortex2} J.W. Fleischer, G. Bartal, O. Cohen, O. Manela, M. Segev,
J. Hudock, and D.N. Christodoulides,
%"Observation of Vortex-Ring "Discrete" Solitons in 2D Photonic Lattices''
Phys. Rev. Lett. \textbf{92}, 123904 (2004).

\bibitem{motihigher} G. Bartal, O. Manela, O. Cohen, J.W. Fleischer, and M. Segev,
%"Observation of Second-Band Vortex Solitons in 2D Photonic Lattices'',
Phys. Rev. Lett. {\bf 95}, 053904 (2005).

\bibitem{MIchristian} C.E. R\"uter, J. Wisniewski, M. Stepi{\'c}, and D. Kip,
Opt. Express {\bf 15}, 6320 (2007).

\bibitem{shand1} M. Stepi{\'c}, E. Smirnov, C.E. R\"uter, L. Pr{\"o}nekke, and D. Kip,
Phys. Rev. {\bf 74}, 046614 (2006).

\bibitem{shand2} E. Smirnov, C.E. R\"uter, M. Stepi{\'c}, D. Kip, and V. Shandarov,
Phys. Rev. E {\bf 74}, 065601 (2006).

\bibitem {brightSoliton1} D. Kip, C.E. R\"uter, R. Dong, Z. Wang, and J. Xu,
%"Higher-band gap soliton formation in defocusing photonic lattices,"
Opt. Lett. \textbf{33}, 2056-2058 (2008).

\bibitem {darksoliton1} R. Dong, C.E. R\"uter, D. Song, J. Xu, and D. Kip,
%"Formation of higher-band dark gap solitons in one dimensional waveguide arrays",
Opt. Express \textbf{18}, 27493-27498 (2010).

\bibitem{Rabi} K. Shandarova, C.E. R\"uter, D. Kip, K.G. Makris, D.N. Christodoulides, O. Peleg, and M. Segev,
%"Experimental observation of Rabi oscillations in photonic lattices",
Phys. Rev. Lett. \textbf{102}, 123905 (2009).

\bibitem{zcv1} Z. Chen, A. Bezryadina, I. Makasyuk, and J. Yang,
Opt. Lett. {\bf 29}, 1656 (2004).

\bibitem{dkv1} R. Vicencio, E. Smirnov, C.E. R\"uter, D. Kip, and M. Stepi{\'c},
Phys. Rev. A {\bf 76}, 033816 (2007).

\bibitem{seg1}
%Z. Chen {\it et al.},
Z. Chen, M. Segev, T.H. Coskun, D.N. Christodoulides, Yu.S. Kivshar, and V.V. Afanasjev,
%Incoherently coupled dark-bright photorefractive solitons
Opt. Lett. {\bf 21}, 1821 (1996).

\bibitem{seg2}
%E.A. Ostrovskaya {\it et al.},
E.A. Ostrovskaya, Yu.S. Kivshar, Z. Chen, and M. Segev,
%Interaction between vector solitons and solitonic gluons.
Opt. Lett. {\bf 24}, 327 (1999).

\bibitem{BA}  Th. Busch and J.R. Anglin,
Phys. Rev. Lett. {\bf 87}, 010401 (2001).

\bibitem{DDB} H.E. Nistazakis, D.J. Frantzeskakis, P.G. Kevrekidis, B.A. Malomed, and R. Carretero-Gonz{\'a}lez,
Phys. Rev. A {\bf 77}, 033612 (2008).

\bibitem{hamburg} C. Becker, S. Stellmer, P. Soltan-Panahi, S. D\"{o}rscher, M. Baumert, E.-M. Richter, J. Kronj\"{a}ger, K. Bongs, and K. Sengstock,
Nature Phys. {\bf 4}, 496 (2008).

\bibitem{engels1} C. Hamner, J.J. Chang, P. Engels, and M.A. Hoefer,
Phys. Rev. Lett. {\bf 106}, 065302 (2011).

\bibitem{engels2} M.A. Hoefer, C. Hamner, J.J. Chang, and P. Engels,
arXiv:1007.4947.

\bibitem{engels3} S. Middelkamp, J.J. Chang, C. Hamner, R. Carretero-Gonz{\'a}lez, P.G. Kevrekidis, V. Achilleos, D.J. Frantzeskakis, P. Schmelcher, and P. Engels,
Phys. Lett. A {\bf 375}, 642 (2011).

\bibitem{rajendran} S. Rajendran, P. Muruganandam, and M. Lakshmanan,
%Interaction of dark–bright solitons in two-component Bose–Einstein condensates.
J. Phys. B {\bf 42}, 145307 (2009).

\bibitem{berloff} C. Yin, N.G. Berloff, V.M. P\'erez-Garc\'{\i}a, V.A. Brazhnyi, and H. Michinel,
%Coherent Atomic Soliton Molecules
arXiv:1003.4617.

\bibitem{VB}
%K.J.H. Law {\it et al.},
K.J.H. Law, P.G. Kevrekidis and L.S. Tuckerman,
%arXiv:1001.4835.
Phys. Rev. Lett. {\bf 105}, 160405 (2010).

\bibitem{aj1} A. {\'A}lvarez, J. Cuevas, F.R. Romero, and P.G. Kevrekidis,
Physica D {\bf 240}, 767 (2011).

\bibitem {WA1} D. Kip,
%"Photorefractive waveguides in oxide crystals: fabrication, properties, and applications,"
Appl. Phys. B \textbf{67}, 131-150 (1998).

\bibitem {bandstructure1} C.E. R\"uter, J. Wisniewski, and D. Kip,
%"Prism coupling method to excite and analyze Floquet-Bloch modes in linear and nonlinear waveguide arrays,"
Opt. Lett. \textbf{31}, 2768-2770 (2006).

\bibitem{interlaced} J. Cuevas, Q.E. Hoq, H. Susanto, and P.G. Kevrekidis,
Physica D {\bf 238}, 2216 (2009).

\end{thebibliography}
\end{document}